\documentclass[11pt]{article}
\usepackage{epsfig,graphicx}

\usepackage{amsmath,amssymb}
\usepackage{amsfonts}

\newcommand{\pd}{\partial}
\newcommand{\Fc}{\mathcal{F}}
\newcommand{\Rc}{\mathcal{R}}

\newcommand{\AP}{\alpha^{\prime}}
\newcommand{\diag}{\mathop{\mathrm{diag}}\nolimits}
\newcommand{\const}{\mathrm{const}}

\begin{document}

\title{Analysis of scalar perturbations in cosmological models with a
non-local scalar field}

\author{Alexey~S.~Koshelev\footnote{Postdoctoral researcher of
FWO-Vlaanderen.}\\
\textit{\small Theoretische Natuurkunde, Vrije Universiteit Brussel
and}\\
\textit{\small The
International Solvay Institutes,}\\
\textit{\small Pleinlaan 2, B-1050 Brussels,
Belgium, alexey.koshelev@vub.ac.be}\\
\\
Sergey~Yu.~Vernov\\
\textit{\small Skobeltsyn Institute of Nuclear Physics, Moscow State
University,}\\
\textit{\small Vorobyevy Gory, 119991, Moscow, Russia,
svernov@theory.sinp.msu.ru}}

\date{~}

\maketitle

\begin{abstract}
We develop the cosmological perturbations formalism in models with a
single non-local scalar field originating from the string field theory
description of the rolling tachyon dynamics. We construct the equation
for the energy density perturbations of the non-local scalar field in
the presence of the arbitrary potential and formulate the local system of
equations for perturbations in the linearized model when both simple and double
roots of the characteristic equation are present. We carry out the general
analysis related to the curvature and entropy
perturbations and consider the most specific
example of perturbations when important quantities in the model become
complex.
\end{abstract}

%



\section{Introduction}

Current cosmological observational data~\cite{data,Komatsu} strongly
support that the present Universe exhibits an accelerated expansion
providing thereby an evidence for a dominating dark energy (DE)
component~\cite{review-de}. Recent results of WMAP~\cite{Komatsu}
together with the data on Ia supernovae and  galaxy clusters
measurements, give the following bounds for the DE state parameter $
w_{\mathrm{DE}}=-1.02^{+0.14}_{-0.16} $. The present cosmological
observations do not exclude a possibility that the DE with $w<-1$
exists, as well as an evolving DE state parameter~$w$. Moreover, the
recent analysis of the observation data indicates that the varying in
time dark energy with the state parameter $w_{DE}$, which crosses the
cosmological constant barrier, gives a better fit than a cosmological
constant (for details see reviews~\cite{Quinmodrev1} and references
therein).

Construction of a stable model with $w<-1$ is a challenge leading to the consideration
of originally stable theories admitting the NEC violation in some limits. Recently a new
class of cosmological models obeying this property which is based on the string field theory
(SFT)~\cite{review-sft} and the $p$-adic string
theory~\cite{padic} has been investigated a lot~\cite{IA1}--\cite{Calcagni2010}.
It is known that both
the SFT and the $p$-adic string theory are UV-complete ones. Thus
one can expect that resulting (effective) models should be free of
pathologies.

Models originating from the SFT are distinguished by presence of
specific non-local operators. The higher derivative terms in principle
may produce the well known Ostrogradski instability~\cite{ostrogradski}
(see also~\cite{PaisU,AV_NEC})\footnote{Additional phantom solutions,
obtained by the Ostrogradski method in some models can be interpreted
as non-physical ones, because the terms with higher-order derivatives
are regarded as corrections essential only at small energies below the
physical cutoff~\cite{Simon, SW}.}. However the Ostrogradski result is
related to higher than two but a finite number of derivatives. In the
case of infinitely many derivatives it is possible that instabilities
do not appear~\cite{noghosts}.

The SFT inspired cosmological models \cite{IA1} are considered as
models for dark energy (DE). The way of solving the Friedmann equations
with a quadratic potential, by reducing them to the Friedmann equations
with many free massive local scalar fields, has been proposed
in~\cite{Koshelev07,AJV0711} (see also~\cite{Vernov2010}). The obtained
local fields satisfy the second order linear differential equations. In
the representation of many scalar fields some of them are normal and
some of them are phantom (ghost) ones~\cite{AJV0711,Vernov2010}.
Cosmological models coming out from the SFT or the $p$-adic string
theory are considered in application to inflation
\cite{Calcagni}--\cite{nongauss} to explain in particular appearance of
non-gaussianities. For a more general discussion on the string
cosmology and coming out of string theory theoretical explanations of
the cosmological observational data the reader is referred to
\cite{string-cosmo}. Other models obeying non-locality and their
cosmological consequences are considered in
\cite{nonlocal,sftnonlocal}.

As a simplest model originating from SFT one can consider a theory with
one scalar field whose kinetic operator is non-local. Equations for
cosmological perturbations in such kind of model where the scalar field
Lagrangian is quadratic covariantly coupled with Einstein gravity were
derived in \cite{KV}. In the present paper we develop and improve that
formalism accounting an arbitrary potential of the scalar field as well as the
presence of double roots of the characteristic equation in the linearized model.
We also carry out the general analysis of curvature and entropy
perturbations and consider the most specific example of perturbations when
characteristic quantities of the model become complex.

The paper is organized as follows. In Section~2 we describe the
non-local non-linear SFT inspired cosmological model. In Section~3
we sketch the construction of background solutions in the
linearized model and perturbation theory for models with non-local
scalar field\footnote{For applications of other multi-field
cosmological models and related technical aspects see for instance
\cite{emails}.}. In Section~4 we consider the perturbations in the
case of two complex conjugate roots.
In Section 5 we consider the case when the analytic function $\Fc(\AP\Box)$,
which defines the kinetic operator in the action, has double roots. In Section~6
we summarize the
obtained results and propose directions for further
investigations.


\section{Model setup}


\subsection{Actions}

We work in $(1+3)$ dimensions, the coordinates are denoted by Greek
indices $\mu,\nu,\dots$ running from 0 to 3. Spatial indexes are
$a,b,\dots$ and they run from 1 to 3. The four-dimensional action
motivated by the string field theory is as follows
\cite{AJK,Koshelev07,AJV0701}:
\begin{equation}
S=\!\int\! d^4x\sqrt{-g}\left(\frac{R}{16\pi G_N}+\frac1{\AP
g_o^2}\left(\frac12
T\Fc_0(\AP\Box)T-{V_{int}(T)}\right)-\Lambda_0\right).
\label{action_model}
\end{equation}
Here $G_N$ is the Newtonian constant: $8\pi G_N=1/M_P^2$, where $M_P$
is the Planck mass, $\AP$ is the string length squared, $g_o$ is the
open string coupling constant.  We use the signature $(-,+,+,+)$,
$g_{\mu\nu}$ is the metric tensor, $R$ is the scalar curvature,
$\Box=D^{\mu}\pd_{\mu}=\frac1{\sqrt{-g}}\partial_{\mu}\sqrt{-g}g^{\mu\nu}\partial_{\nu}$
 and $D_\mu$ being a covariant
derivative, $T$ is a scalar field primarily associated with the open
string tachyon. The function $\Fc_0(\AP\Box)$ may be not a polynomial
manifestly producing thereby the non-locality. Fields are dimensionless
while $[g_o]=\mathrm{length}$. The potential $V_{int}(T)$, which is an open
string tachyon
self-interaction, does not have a quadratic term. It is convenient to
introduce dimensionless
coordinates $\bar{x}_\mu=x_\mu/\sqrt{\alpha^{\prime}}$,
the dimensionless gravitational constant $\overline{G}_N=G_N/\alpha^{\prime}$, and
the dimensionless coupling
constant $\bar g_{o}=g_{o}/\sqrt{\alpha^{\prime}}$. In the following formulae we always use dimensionless
coordinates and parameters omitting bars over them.

Function $\Fc_0$ is assumed to be an analytic function of its argument, such that one can represent it by the convergent series
expansion with real coefficients:
\begin{equation}
\Fc_0=\sum\limits_{n=0}^{\infty}f_n\Box^n\quad\mbox{ and }\quad
f_n\in\mathbb{R}.
\end{equation}
Equations of motion are
\begin{eqnarray}
G_{\mu\nu}=\frac{8\pi G_N}{g_o^2}T_{\mu\nu},
&&\label{EOJ_g}\\
\Fc_0(\Box)T=V_{int}'(T),&& \label{EOJ_tau}
\end{eqnarray}
where $G_{\mu\nu}$ is the Einstein tensor, the energy--momentum (stress)
tensor is as follows
\begin{eqnarray}
T_{\mu\nu}&=&\frac{1}{2}\sum_{n=1}^\infty
f_n\sum_{l=0}^{n-1}\Bigl(\pd_\mu\Box^l T \pd_\nu\Box^{n-1-l} T
+\pd_\nu\Box^l T \pd_\mu\Box^{n-1-l} T -{}\nonumber \\
&&{}-g_{\mu\nu}\left(g^{\rho\sigma}
\pd_\rho\Box^l T \pd_\sigma\Box^{n-1-l} T +\Box^l T \Box^{n-l} T \right)\Bigr)-{}\nonumber\\
&& {}-g_{\mu\nu}\left(g_o^2\Lambda_0-
\left(\frac12T\Fc_0(\Box)T-V_{int}(T)\right)\right).\nonumber
\end{eqnarray}
It is easy to check that the Bianchi identity is satisfied on-shell and
for $\Fc_0=f_1\Box+f_0$ the usual energy--momentum tensor for the
massive scalar field is reproduced.

Let us emphasize that the potential of the field $T$ is
$V=-\frac{f_0}{2}T^2+V_{int}(T)$. Let $T_0$ be an extremum of the
potential $V$. One can linearize the theory in its neighborhood
using $T=T_0+\tau$.  To second order in $\tau$  one gets the following
action
\begin{equation}
S_2=\int d^4x\sqrt{-g}\left(\frac{R}{16\pi
G_N}+\frac1{2g_o^2}\tau\Fc(\Box)\tau-\Lambda\right),
\label{action_model2}
\end{equation}
where $\Fc=\Fc_0-V(T_0)''$ and
$\Lambda=\Lambda_0+\frac{V(T_0)}{g_o^2}$.  Equations
(\ref{EOJ_g}) and (\ref{EOJ_tau}) are valid for the latter action after
the replacement $\Fc_0\to\Fc$ and $\Lambda_0\to\Lambda$ at $V_{int}(T)=0$.
Note that all Taylor series coefficients $f_n$, except $f_0$, are the same
for $\Fc_0$ and $\Fc$. Equation (\ref{EOJ_tau}) now is
\begin{equation}
\Fc(\Box)\tau=0.  \label{EOJ_tau_loc}
\end{equation}
Non-local cosmological models of type (\ref{action_model2}) with
\begin{equation*}
\label{F_SFT} \Fc_{\mathrm{sft}}(\Box)={}-\xi^2 \Box+1-c\:e^{-2\Box},
\end{equation*}
were previously considered
in~\cite{AJV0701,AJV0711,MulrunyNunes}\footnote{In~\cite{MulrunyNunes}
for example it has been shown that solving the non-local equations
using the localization technique is fully equivalent to
reformulating the problem using the diffusion-like partial
differential equations. One can fix the initial data for the
partial differential equation, using the initial data of the
special local fields. This specifies initial data for a non-linear
model, and these initial data can be (numerically) evolved into
the full non-linear regime using the diffusion-like partial
differential equation.}. Actions (\ref{action_model}) and
(\ref{action_model2}) are of the main concern of the present
paper.


\subsection{Background solutions construction in the linearized mo\-del}

While solution construction in the full non-linear model
(\ref{action_model}) is not yet known the classical solutions to
equations coming out the linearized action (\ref{action_model2})
were studied and analyzed
in~\cite{Koshelev07,AJV0701,AJV0711,MulrunyNunes,BarnabyKamran2,KV,Vernov2010}.
Thus, we just briefly notice the key points useful for purposes of
the present paper.

The main idea of finding solutions to the equations of motion is
to start with equation (\ref{EOJ_tau_loc}) and to solve it,
assuming the function $\tau$ is  a sum of eigenfunctions of the
d'Alembert operator:
\begin{equation}
\tau=\!\sum\limits_i\tau_i,~\mbox{where}~\Box\tau_i=J_i\tau_i~\mbox{
and }~\Fc(J_i)=0~\mbox{ for any }~i=1,\dots,N.
 \label{tau_sum}
\end{equation}
Hereafter we use $N$ (which can be infinite as well) denoting the
number of roots and omit it in writing explicit summation
limits over $i$. Without loss of generality we assume that for any
$i_1$ and $i_2\neq i_1$ condition $J_{i_1}\neq J_{i_2}$ is
satisfied.  In this way of solving all the information is
extracted from the roots of the \textit{characteristic}
equation~$\Fc(J)=0$.
We can consider the solution $\tau$  as a general solution if all
roots of $\Fc$ are simple. The analysis is more complicated in the
case of double roots~\cite{Vernov2010}. We consider this
case separately in Section 5.

In an arbitrary metric the energy--momentum tensor in (\ref{EOJ_g})
evaluated on such a solution is~\cite{KV}
\begin{equation}
T_{\mu\nu}=\sum_i\Fc'(J_i)\left(\pd_\mu\tau_i\pd_\nu\tau_i
-\frac{g_{\mu\nu}}{2}\left(g^{\rho\sigma}\pd_\rho\tau_i\pd_\sigma\tau_i+J_i\tau_i^2\right)
\right)-g_o^2g_{\mu\nu}\Lambda. \label{EOJ_g_onshell}
\end{equation}
The last formula is exactly the energy--momentum tensor of many
free massive scalar fields. If $\Fc(J)$ has simple real roots,
then positive and negative values of $\Fc'(J_i)$ alternate, so we
can obtain phantom fields.
 Using formula (\ref{EOJ_g_onshell}) we
obtain the Ostrogradski representation~\cite{ostrogradski,PaisU}
for action (\ref{action_model2}):
\begin{equation*}
S_3=\! \int\! d^4x\sqrt{-g}\left(\frac{R}{16\pi G_N}-\Lambda-
\sum_{i=1}\frac{\Fc'(J_i)}{2g_o^2}\left(g^{\mu\nu}\pd_\mu\tau_i\pd_\nu\tau_i
+J_i\tau_i^2\right)\right).
\end{equation*}
one can see that $S_3$ is a local action if the number of roots $N$ is finite.


\subsection{Application to Friedmann--Robertson--Walker Universe}

We stress that all the above formulae are valid for an arbitrary
metric and the general solution. From now on, however, the only
metric we will be interested in is the spatially flat
Friedmann--Robertson--Walker (FRW) metric with
the interval given by
\begin{equation}
\label{mFr} ds^2={}-dt^2+a^2(t)\left(dx_1^2+dx_2^2+dx_3^2\right)
\end{equation}
where $a(t)$ is the scale factor, $t$ is the cosmic time.


 Background solutions
for $\tau$ are taken to be space-homogeneous. The
energy--momentum tensor in (\ref{EOJ_g}) in this metric can be written
in the form of a perfect fluid $T^\mu_\nu=\diag(-\varrho,p,p,p)$, where
\begin{equation}
\begin{array}{rcl}
\displaystyle \varrho&=&\displaystyle\frac1{2}\sum_{n=1}^\infty
f_n\sum_{l=0}^{n-1} \Bigl(\pd_t\Box^l T \pd_t\Box^{n-1-l} T +
\Box^l T \Box^{n-l} T \Bigr)-{}\\[2.7mm]&&\displaystyle {}-
\frac12T\Fc_0(\Box)T+V_{int}(T)+g_o^2\Lambda_0,\\[2.7mm]
\displaystyle p&=&\displaystyle\frac1{2}\sum_{n=1}^\infty
f_n\sum_{l=0}^{n-1}\left(\pd_t\Box^l T \pd_t\Box^{n-1-l} T -
\Box^l T \Box^{n-l} T
\right)+{}\\[2.7mm]&&\displaystyle{}+\frac12 T\Fc_0(\Box)T-V_{int}(T)-g_o^2\Lambda_0.
\end{array}\label{ep_sol}
\end{equation}
Obviously we can rewrite equations (\ref{EOJ_g}) in the
canonical form:
\begin{equation}
3H^2={8\pi G}\varrho,\qquad \dot H={}-{4\pi G}(\varrho+p), \label{FrEOM}
\end{equation}
where $G\equiv G_N/g_o^2$ is a dimensionless analog of the Newtonian
constant, $H=\dot a/a$ and a dot
denotes a derivative with respect to the cosmic time $t$. The
consequence of (\ref{FrEOM}) is the conservation equation:
\begin{equation}
\dot\varrho+3H(\varrho+p)=0. \label{EQUrho}
\end{equation}
which corresponds to the non-local field equation (\ref{EOJ_tau}).

Some progress is possible in the linearized model when the metric is fixed
to be of the FRW type. The individual equations in (\ref{tau_sum}) in this
metric read
\begin{equation}
\label{FrEOMtaucsf_frw} \ddot\tau_i+3H\dot\tau_i+J_i\tau_i=0
\end{equation}
The full system of equations of motion has the
fixed points at $\tau_i=0$ and $3H^2=3H_0^2=8\pi G_N\Lambda$,
which is real at $\Lambda>0$. Equations
(\ref{FrEOMtaucsf_frw}) together with Friedmann equations describe the late time
evolution of the model with
Lagrangian (\ref{action_model}). This model possesses an approximate solution
with all
scalar fields tending to the minimum of the potential (i.e. $\tau_i\to0$)
and the Hubble parameter going to the constant. Such solution was
constructed numerically and was proven to be a solution in
\cite{Jukovskaya0707}. Also the asymptotic form of this solution was
derived in \cite{Koshelev07}.

It is instructive to investigate the Lyapunov stability of
the fixed point. Using formulae from~\cite{ABV_TMF2010}, we come to
conclusion the fixed point is asymptotically stable at
\begin{equation}
H_0>0, \qquad \Re e(J_i)<0.\label{Stabcond_frw}
\end{equation}
If at the fixed point $H_0<0$ or $\Re e(J_i)>0$ for some $i$, then this fixed
point is
unstable. At $\Re e(J_i)=0$ for some $i$ or $H_0=0$ one can not use the Lyapunov
theorem to
analyse the stability of the fixed point. Note that the conditions
(\ref{Stabcond_frw}) are sufficient for stability in not only the
Friedmann--Robertson--Walker metric but also the Bianchi I
metric~\cite{ABV_TMF2010}. In this paper we shall extend this analysis
and consider the stability of
the fixed point with respect to arbitrary perturbations.

To compute an approximate solution to (\ref{FrEOMtaucsf_frw}) one starts
with a constant $H=H_0$ and then computes the correction to $H$ using
Friedmann equations. Then the procedure can be iterated to compute higher
corrections. It was proven in \cite{Koshelev07} such
iteration does converge.

Solution to (\ref{FrEOMtaucsf_frw}) with constant $H=H_0$ is obviously
\begin{equation}
\label{FrEOMtaucsfsoltau_frw}
\tau_i=\tau_{i+}e^{\alpha_{i+}t}+\tau_{i-}e^{\alpha_{i-}t}
\end{equation}
where
$\alpha_{i\pm}=\frac{3H_0}{2}\left(-1\pm\sqrt{1-\frac{4J_i}{9H_0^2}}\right)$.
Considering $\tau_i$ we see that only one term in the solution converges
when $t\to\infty$ in general (if both converge we select the slowest
one). Let's assume it is the first one proportional to $\tau_{+}$
constant. Then in order to pick the (slowest) convergent solution we
put $\tau_{-}=0$.\footnote{Hereafter we adopt the rule
$\sqrt{z^*}=\sqrt{z}^*$ meaning that the phase of the complex number
runs in the interval $[-\pi,\pi)$ and for $z=re^{i\sigma}$ the square
root is $|\sqrt{r}|e^{i\sigma/2}$.}

The first correction to the constant Hubble parameter in case only decaying
modes in $\tau_i$ are present is
\begin{equation}
\label{FrEOMtaucsfsolH_frw}
H=H_0+h=H_0+h_0\sum\tau_i^2.
\end{equation}
Constant $h_0$ is not independent and is related with $\tau_{i+}$. We note
that $h$ is of order $\tau_i^2$. The last expression is a good
approximation for $H$ in the asymptotic regime when $h\ll H_0$.
Further one can find the scale factor to be
\begin{equation}
\label{rhop10csola_frw}
a=a_0\exp\left(H_0t+\frac{h_0}2\sum\frac{\tau_i^2}{\alpha_{i+}}\right).
\end{equation}


\section{Cosmological perturbations with single \textit{non}-local scalar field}

\subsection{General analysis}

Now we turn to the main problem of the present paper: derivation of
cosmological perturbation equations in models with a non-local scalar
field. We are focused on the scalar perturbations, because both vector
and tensor perturbations exhibit no instabilities~\cite{Mukhanov}.
Scalar metric perturbations are given by four arbitrary scalar
functions~\cite{bardeen,Mukhanov}. Changing the coordinate system one
can both produce fictitious perturbations and remove real ones. Natural
way to distinct real and fictitious perturbations is introducing
gauge-invariant variables, which are free of these complications and
are equal to zero for a system without perturbations.  There exist two
independent gauge-invariant variables (the Bardeen potentials), which
fully determine the scalar perturbations of the metric
tensor~\cite{bardeen,Mukhanov,hwangnoh,addon,KV}. To construct the
perturbation equations one can use the longitudinal
(conformal-Newtonian) gauge, in which the interval (\ref{mFr}) with
scalar perturbations has the following form (in terms of the Bardeen
potentials):
\begin{equation}
ds^2=a(\eta)^2\left(-(1+2\Phi)d\eta^2+\delta_{ab}^{(3)}(1-2\Psi)dx^adx^b\right)
\label{deltametric}
\end{equation}
where $\eta$ is the conformal time related to the cosmic one as
$a(\eta)d\eta=dt$. The the Bardeen potentials $\Phi$ and $\Psi$ are as
usually Fourier transformed with respect to the spatial coordinates
$x^a$ having thereby the following form:
$\Phi(\eta,x^a)=\Phi(\eta,k)e^{ik_ax^a}$ and similar for $\Psi$. The
obtained equations contain only gauge invariant variables, so they are
valid in an arbitrary gauge.

Although the metric perturbations are defined in the conformal time
frame in the sequel the cosmic time $t$ will be used as the function
argument and all the equations will be formulated with $t$ as the
evolution parameter.

To the background order energy density and pressure  are given by
(\ref{ep_sol}). To the perturbed order one has
\begin{eqnarray}
\delta\!\varrho&=&\frac1{2}\sum_{n=1}^\infty
f_n\sum_{l=0}^{n-1}\left(\pd_t\delta(\Box^l T )\pd_t\Box^{n-1-l} T
+\pd_t\Box^l T \pd_t\delta(\Box^{n-1-l} T )-{}\right.\nonumber\\
&&{}-2\Phi\pd_t\Box^l T \pd_t\Box^{n-1-l} T +{}\label{drho}\\
&&{}+\left.\delta(\Box^l T )\Box^{n-l} T
+\Box^l T \delta(\Box^{n-l} T )\right)-\frac1{2g_o^2}(TV_{int}''-V_{int}')\delta T\nonumber,\\
\delta\! p&=&\frac1{2}\sum_{n=1}^\infty
f_n\sum_{l=0}^{n-1}\left(\pd_t\delta(\Box^l T )\pd_t\Box^{n-1-l} T
+\pd_t\Box^l T \pd_t\delta(\Box^{n-1-l} T )-{}\nonumber\right.\\
&&{}-2\Phi\pd_t\Box^l T \pd_t\Box^{n-1-l} T -{}
\label{dp}\\&&{}-\left.\delta(\Box^l T )\Box^{n-l} T -\Box^l T
\delta(\Box^{n-l} T )\right)+\frac1{2g_o^2}(TV_{int}''-V_{int}')\delta T,\nonumber\\
v^s&=&\frac k {a(\varrho+p)}\sum_{n=1}^\infty f_n\sum_{l=0}^{n-1}
\pd_t\Box^l T \delta\left(\Box^{n-1-l} T \right)\label{v},\\
\pi^s&=&0.
\end{eqnarray}
where $v^s$ gives the perturbed $T^0_a$ components of the
stress-energy tensor and $\pi^s$ is the anisotropic stress. Using
the Einstein equations one gets that $\pi^s=0$ is equivalent to
$\Phi=\Psi$. The Bardeen potential $\Psi$ is proportional to the
gauge invariant total energy perturbation
\begin{equation}
\varepsilon\equiv\frac{\delta\varrho}{\varrho}+3(1+\omega)Hv^s\frac{a}{k}={}
-\frac{k^2}{4\pi
G\varrho a^2}\Psi.
\label{bardeenpsi}
\end{equation}
The function $\varepsilon$ is a solution of the following linear
differential equation (see details in~\cite{KV}):
\begin{equation}
\begin{array}{l}
\displaystyle\ddot\varepsilon+H\left(2+3c_s^2-6w\right)\dot\varepsilon +{}\\
\displaystyle{}+\left(\dot
H(1-3w)-15H^2w+9H^2c_s^2+\frac{k^2}{a^2}\right)\varepsilon+\frac{k^2}{a^2\varrho}\Delta=0.
\end{array}
\label{deltaGIeps01nlscalar}
\end{equation}
Here $w=p/\rho$ is the equation of  state parameter, $c_s^2=\dot
p/\dot\rho$ is the speed of sound,  $k=\sqrt{k_ak^a}$ is the comoving
wavenumber and
\begin{equation*}
\begin{array}{l}
\displaystyle\Delta=\delta p-\delta\varrho+(1-c_s^2)\frac{a}{k}\dot\varrho v^s=\frac {(1-c_s^2)\dot\varrho}{\varrho+p}
\sum\limits_{n=1}^\infty f_n\sum\limits_{l=0}^{n-1} \pd_t\Box^l T
\delta(\Box^{n-1-l}T)-{}\\ \displaystyle
{}-\!\sum\limits_{n=1}^\infty
f_n\!\sum\limits_{l=0}^{n-1}\left(\delta(\Box^l T )\Box^{n-l} T
+\Box^l T \delta(\Box^{n-l} T )\right)+\frac1{g_o^2}(TV_{int}''-V_{int}')\delta T.
\end{array}
\end{equation*}
The latter quantity is identically zero for a local scalar field,
i.e. in the case $\Fc(\Box)=f_1\Box+f_0$.
Therefore, $\Delta\neq 0$ is the attribute of the non-locality here.

For the linearized model (\ref{action_model2}) we can consider the
background solution as given by (\ref{tau_sum}) to obtain $\Delta$ in
the more convenient form. To do this the following relation is  useful
\begin{equation}
\label{deltabox1}
\delta(\Box^n\tau)=\Box^n\delta\!\tau+\sum_{m=0}^{n-1}\Box^m(\delta\Box)\Box^{n-1-m}\tau.
\end{equation}
Using (\ref{tau_sum}) and the well-known formula
\begin{equation*}
\sum_{m=0}^{n-1}x^m=\frac{x^n-1}{x-1},
\end{equation*}
 one has
\begin{equation}
\label{deltabox2}
\delta(\Box^n\tau)=\Box^n\delta\!\tau+\sum_i\frac{\Box^n-J^n_i}{\Box-J_i}(\delta\Box)\tau_i.
\end{equation}

Perturbing the equation of motion for $\tau$, one has
\begin{equation}
\delta(\Fc\tau)=\sum_{n=0}^\infty f_n\delta(\Box^n\tau)=0.\label{dtau}
\end{equation}
More explicitly this equation can be written as
\begin{equation}
\delta(\Fc\tau)=\Fc\sum_i\left(\frac1{\Box-J_i}(\delta\Box)\tau_i+\delta\!\tau_i\right)=0
\label{dtauexplicit}
\end{equation}
where we have put $\delta\!\tau=\sum\limits_i\delta\!\tau_i$.

It follows from (\ref{EOJ_g_onshell}) that if for
some $J_k$ we have $\tau_k=0$ as a background solution,
then $\delta\!\tau_k$, contributes only to the second order in the
energy--momentum tensor perturbations. In this paper we consider perturbations
only to the first order, and therefore for all $\tau_k=0$   we can put
$\delta\!\tau_k=0$ without loss of generality. If $\Fc$ has an infinite number
of roots, but we select as a background the function
 $\tau$, which includes only a  finite number of $\tau_k$, then only a finite
number of perturbations   $\delta\!\tau_k$ give contribution to the first
 order perturbation equations, whereas in the second order all perturbations are
important.

After some algebra one can get the following expression for $\Delta$
\begin{equation}
\Delta={}-\frac{2}{\varrho+p}\sum_{m,l}\Fc'(J_m)\Fc'(J_l)J_m\tau_m\dot\tau_m\dot\tau_l^2\zeta_{ml}\,,
\label{explicitDelta}
\end{equation}
where
$\zeta_{ij}=\frac{\delta\tau_i}{\dot\tau_i}-\frac{\delta\tau_j}{\dot\tau_j}$
satisfy the following set of equations
\begin{equation}
\begin{array}{l}
\displaystyle
\ddot{\zeta}_{ij}+\left(3H+\frac{\ddot\tau_{i}}{\dot\tau_i}+
\frac{\ddot\tau_{j}}{\dot\tau_j}\right)\dot\zeta_{ij}+\left(-3\dot H+\frac{k^2}{a^2}\right)\zeta_{ij}=\\
\displaystyle=
\left(\frac{J_i\tau_{i}}{\dot\tau_i}-\frac{J_j\tau_{j}}{\dot\tau_j}\right)
\left(\sum_m\frac{\Fc'(J_m)\dot\tau_m^2}{\varrho+p}\left(\dot\zeta_{im}+
\dot\zeta_{jm}\right)+\frac2{1+w}\varepsilon\right).
\end{array}
\label{deltaijeps}
\end{equation}
Equation (\ref{deltaGIeps01nlscalar}) with the above derived $\Delta$
and equations (\ref{deltaijeps}) in a closed form describe the
perturbations in the case of linearized model. Comparing these equation
with the equation for perturbations in a system with many fields (see
eqs. (82) and (85) in~\cite{hwangnoh}) we see they do coincide. Thus
perturbations become equivalent in the model with one free non-local
scalar field and in the model with many local scalar fields. In our
model, however, the quantity which should be considered as energy
density perturbation is $\varepsilon$. Functions $\zeta_{ij}$ play
auxiliary role and normally should not be given an interpretation.


\subsection{Curvature and entropy perturbations}

To understand better what is going on in the discussed
type of models it is instructive to see how the curvature and entropy
perturbations behave in our model having infinitely many scalar degrees of
freedom (interesting results in  analysing this parameters in case of many
fields can be found for instance in \cite{manyref}).


Comoving curvature perturbations can be expressed as
\begin{equation}
  \Rc=\Psi-\frac{H}{\dot H}(\dot\Psi+H\Psi)\label{isocurv}
\end{equation}
where $\Psi$ is the Bardeen potential connected in turn with $\varepsilon$ as
(\ref{bardeenpsi}).
Entropy perturbations defined as $e=\delta p-c_s^2\delta\varrho$ can be found as
\begin{equation}
\frac e{\varrho}=\varepsilon-(1+c_s^2)\frac{a^2}{k^2}\Delta
  \label{entropy}
\end{equation}
Both quantities are gauge invariant and play crucial role in computing
various spectral indexes.

In order to figure out the behavior for the curvature and entropy
perturbations it is enough to
find out the behavior of only two functions: $\varepsilon$ and $\Delta$. Such a
generic form in formulae
(\ref{isocurv}) and (\ref{entropy}) persists because the theory can be written
in the local form. It is nevertheless obvious
that it would be difficult to reach similar results in a general
case, when $V_{int}\neq 0$ and equation (\ref{EOJ_g}) is not linear in $T$.
Moreover, it is not transparent that one can make a significant progress for a
general operator $\Fc$ even in the linearized model.

There are several specific situations in which we can outline the strategy of
the further analysis.
\begin{itemize}
\item The first one is already mentioned above and refers to the background
configuration where only the finite number of scalar fields is excited. Then in
the linear perturbations we will have no impact of the perturbation modes
related to the zero background fields. This statement is obvious having in mind
that the local action is quadratic and diagonal in scalar fields.
\item The second situation corresponds to particular form of the function $\Fc$
such
that its roots form a sequence $J_i$ and one can arrange them such that
$J_i-J_{i+1}\to0$ when $i$ tends to infinity. Then in the asymptotic regime
where solutions
for the scalar fields are dumped plane waves (\ref{FrEOMtaucsfsoltau_frw}) one
gets the factor in the RHS of
equation (\ref{deltaijeps}) tending to zero and equations for corresponding
$\zeta_{ii+1}$ become homogeneous. In this case one can easily solve this
equation which now reads
\begin{equation}
\ddot{\zeta}_{ii+1}+c\dot\zeta_{ii+1}+\left(-3\dot
H+\frac{k^2}{a^2}\right)\zeta_{ii+1}=0.
\label{deltaijepshomo}
\end{equation}
Here we assumed the coefficient in front of the first derivative to be
a constant while the background solutions are the planewaves. Two modes come
out when the time grows
\begin{equation}
{\zeta}_{ii+1}\approx C_1+C_2e^{-ct}
\label{deltaijepshomomodes}
\end{equation}
and therefore we see that there is a decaying or growing mode depending on the
sign of $c$. This constant in the asymptotic regime is given by
\begin{equation}
c=3H_0-\left(\sqrt{J_i}+\sqrt{J_{i+1}}\right).
\label{deltaijepshomomodesconst}
\end{equation}
Recall that the stability of the asymptotic solution requires $\Re e(J_i)<0$
(see (\ref{Stabcond_frw})) and it guarantees that $c>0$ and perturbations vanish.
Even though it is not the final answer and one still has to solve coupled
equations the claim is that only a finite number of $\zeta_{ij}$ are really
coupled with $\varepsilon$ while other functions $\zeta_{ij}$ just produce the
inhomogeneity in equation for $\varepsilon$ (\ref{deltaGIeps01nlscalar}).
Mathematically we can put all this decoupled modes to be zero but than the
problem becomes completely equivalent to the previous case with a finite number
of fields in the play. Here, however, we keep trace of other fields even though
the are effectively decoupled from the system of equations.

One example of the operator giving such a behavior is
\begin{equation}
\Fc(J)=\alpha-e^{\beta J}
  \label{exp}
\end{equation}
Roots are given by
\begin{equation}
J_k=\frac1{\beta}(\log(\alpha)+2\pi i k)
  \label{exproots}
\end{equation}
Resummation in $\Delta$ (\ref{explicitDelta}) for all the components which
are effectively decoupled (assuming this happens from some $k=k_0$) gives
\begin{equation}
\Delta_0=-\frac{2C_2}{\varrho+p}\frac{e^{-ct}}{1+e^{-ct}}
\label{deltaDec}
\end{equation}
where $C_2$ and $c$ are from eq. (\ref{deltaijepshomomodesconst}) when $i=k_0$.

\item The third situation is when the values of $\sqrt{J_i}$ are equidistant
meaning $\sqrt{J_{i+1}}-\sqrt{J_i}=c_J$. Then equation (\ref{deltaijeps})
in the asymptotic regime becomes
\begin{equation}
\begin{array}{l}
\displaystyle
\ddot{\zeta}_{i+1i}+\left(3H+2\sqrt{J_i}+c_J
\right)\dot\zeta_{i+1i}+\left(-3\dot
H+\frac{k^2}{a^2}\right)\zeta_{i+1i}=\\
\displaystyle=
c_J
\left(\sum_m\frac{\Fc'(J_m)\dot\tau_m^2}{\varrho+p}\left(\dot\zeta_{i+1m}+
\dot\zeta_{im}\right)+\frac2{1+w}\varepsilon\right).
\end{array}
\label{deltaijepsconst}
\end{equation}
The advantage with respect to the general situation is that the the
inhomogeneous part of this equation is universal for any number $i$ meaning
that we can construct homogeneous equations consdering the difference of the
latter equations for some $i$ and $j$.
\end{itemize}
It is important to say that all these cases even giving some insight into the
problem are not very simple to analyze. We hope to address this issue in the
future analysis of models of this type.


\section{Complex roots $J$ in the linearized model}


\subsection{One pair of complex conjugate roots $J_1=J$ and $J_2=J^*$. The background}

If a complex number $J$ is a root of $\Fc$, then
$J^*$ is a root of $\Fc$ as well.  System (\ref{FrEOM}) becomes:
\begin{equation}
\label{FrEOMtauc}
\begin{array}{l}
\displaystyle H^2=\frac{8}{3}\pi
G_N\!\left[\frac{\Fc'(J)}{2g_o}\left(\dot\tau^2
+J\tau^2\right)+\frac{\Fc'({J^*})}{2g_o}\!\left({\dot{\tau^*}}^2
+{J^*}{\tau^*}^2\right)+\Lambda\right],\\[2.7mm]
\displaystyle \dot H={}-\frac{4\pi G_N}{g_o}
\left[\Fc'(J)\dot{\tau}^2+\Fc'({J^*}){\dot{\tau^*}}^2\right].
\end{array}
\end{equation}
In terms of real fields $\phi$ and $\psi$ such that
$
\tau=\phi+i\psi$, $\tau^*=\phi-i\psi$,
we get the following kinetic term:
\begin{equation}
E_k=\frac{\Fc'(J)}{2g_o}\dot\tau^2
+\frac{\Fc'({J^*})}{2g_o}{\dot{\tau^*}}^2=
\frac{d_r}{g_o}\left(\dot\phi^2-\dot
\psi^2\right)+2\frac{d_i}{g_o}\dot\phi\dot\psi,
\end{equation}
where $d_r=\Re e(\Fc'(J))$ and $d_i=\Im m(\Fc'(J))$. In the case
$d_i\neq 0$ $E_k$ has  a nondiagonal form. To diagonalize kinetic term
we make the following transformation:
\begin{equation*}
\chi=\phi+C_1\psi, \quad \nu={}-C_1\phi+\psi,\quad C_1\equiv\frac{d_r+\sqrt{d_r^2+d_i^2}}{\sqrt{d_i}}.
\end{equation*}

In terms of $\chi$ and $\nu$ system (\ref{FrEOMtauc}) has the following
form:
\begin{equation}
\label{FrEOMtaucReal}
\begin{array}{l}
\displaystyle H^2=\frac{8}{3}\pi G_N\left[\frac{C}{2g_o^2}
\left(\dot\nu^2-\dot\xi^2+J_r(\nu^2-\xi^2)+2J_m\nu\xi\right)+\Lambda\right],\\
\displaystyle \dot H=\frac{4\pi G_N C}{g_o^2}\left(\dot\chi^2
-\dot\nu^2\right),
\end{array}
\end{equation}
where $J_r=\Re e(J)$, $\ J_m=\Im m(J)$, $C=\frac{
d_i^2(d_r^2+d_i^2)\left(d_r+\sqrt{d_r^2+d_i^2}\right)}{d_r^2+d_i^2+d_r\sqrt{d_r^2+d_i^2}}$.
So, in the case of two complex conjugated roots we get a quintom model
(for details of quintom models see reviews~\cite{Quinmodrev1}).

What is interesting (but not surprising, though) one cannot have
non-inter\-acting fields passing to the real components. Precisely,
fields will be quadratically coupled in the Lagrangian. It means that
the usual intuition about field properties (like signs of coefficients
in front the kinetic term or the mass term) may not work.

Following the method outlined in Section 2.3 we find the asymptotic
solution for the scalar fields with constant
$H=H_0$ to be
\begin{equation}
\label{FrEOMtaucsfsoltau}
\tau=\tau_{+}e^{\alpha_+t}+\tau_{-}e^{\alpha_-t},\qquad
\tau^*=\tau^*_{+}e^{\alpha_+^*t}+\tau^*_{-}e^{\alpha_-^*t}
\end{equation}
where
$\alpha_{\pm}=\frac{3H_0}{2}\left(-1\pm\sqrt{1-\frac{4J}{9H_0^2}}\right)$.
We assume the first term proportional to $\tau_{+}$
does converge and put $\tau_{-}=0$. Further we define $\tau_{+}\equiv\tau_0$
and $\alpha_+\equiv\alpha$.

The first correction to the constant Hubble parameter and to the scale factor in
case only decaying modes in $\tau$ are present gives
\begin{equation}
\label{FrEOMtaucsfsolH}
H=H_0+h=H_0+h_0\left(\tau^2+{\tau^*}^2\right).
\end{equation}
and
\begin{equation}
\label{rhop10csola}
a=a_0\exp\left(H_0t+\frac{h_0}2\left(\frac{\tau^2}{\alpha}+\frac{{\tau^*}^2}{\alpha^*}\right)\right).
\end{equation}


\subsection{Cosmological perturbations in the neighborhood of the solution with complex masses}

Configurations with a single scalar field were widely studied and those
appearing in the non-local models do not have any distinguished
properties. Roughly speaking configurations with many scalar fields
were explored as well but we have here new models featuring complex
masses and complex coefficients in  front of the kinetic terms. As it
was stressed above there is no problem with this for the physics of our
models while properties of such models, in particular the cosmological
perturbations with such scalar fields were not studied in depth. Thus
we focus on perturbations in the configuration with complex roots $J$.
The simplest case is one pair of complex conjugate roots where the
background quantities were derived in previous Subsection.


First we note that the only function
$\zeta_{ij}$  is $\zeta_{12}$ which we shall denote $\zeta$. Thus, there are only two equations in the
system. We focus on the asymptotic regime $h\ll H_0$ and after some algebra one arrives
to the following system of equations
\begin{equation}
\begin{array}{l}
\displaystyle
(\varrho+p)\left(\ddot{\zeta}+\left(3H_0+\alpha+\alpha^*\right)\dot\zeta
+\left(-3\dot H+\frac{k^2}{a_0^2}e^{-2H_0t}\right)\zeta\right)=\\
\displaystyle =\left(\frac{J}{\alpha}-\frac{{J^*}}{\alpha^*}\right)
\left(\left[{\Fc'(J^*){\alpha^*}^2{\tau ^*}^2}-{\Fc'(J)\alpha^2\tau ^2}
\right]\dot\zeta+{2g_o^2\Lambda}\varepsilon\right),
\end{array}
\label{deltaijepsex}
\end{equation}
\begin{equation}
\begin{array}{l}
\displaystyle \ddot\varepsilon+\dot\varepsilon H_0(8+3c_s^2)
+\varepsilon\left(15H_0^2+9H_0^2c_s^2+\frac{k^2}{a_0^2}e^{-2H_0t}\right)={}\\
\displaystyle
{}=\frac{2k^2\Fc'(J)\Fc'(J^*)\alpha^2{\alpha^*}^2\tau_0^2{\tau_0^*}^2}
{(\varrho+p)a_0^2g_o^2\Lambda}\left(\frac{J}{\alpha}
-\frac{{J^*}}{\alpha^*}\right){e^{2(-H_0+\alpha+\alpha^*)t}}\zeta.
\end{array}
\label{deltaepsijex}
\end{equation}
where we should use
\begin{equation*}
\begin{array}{ll}
\displaystyle \dot H&\displaystyle =2h_0\left(\tau ^2\alpha+{\tau
^*}^2\alpha^*\right),\\ \displaystyle \rho+p&\displaystyle =\Fc'(J)\tau
^2\alpha^2+\Fc'({J^*}){\tau
^*}^2{\alpha^*}^2,\\
\displaystyle c_s^2&\displaystyle =\frac{\Fc'(J)\alpha\tau
^2\left(\alpha^2 -J\right)+\Fc'({J^*})\alpha^*{\tau
^*}^2\left({\alpha^*}^2 -{J^*}\right)}{\Fc'(J)\alpha\tau
^2\left(\alpha^2 +J\right)+\Fc'({J^*})\alpha^*{\tau
^*}^2\left({\alpha^*}^2 +{J^*}\right)}.
\end{array}
\end{equation*}
The latter system of equations is ready to be solved numerically but in
order to get some insight in what is going on it is instructive to make
some assumptions about the value $J$. This makes some analytic progress
possible.

We recall the SFT origin of the model. Practically this means that
values of $J$ are determined with the string scales while $H_0$ is
expected to be much smaller. Therefore, it is natural to assume that
$|\sqrt{J}|\gg H_0$. This implies $\alpha\approx i\sqrt{J}$.
Using the explicit expression for $\tau_1=\tau_0e^{\alpha
t}\approx\tau_0e^{i\sqrt{J}t}$, representing $\alpha=x/2+iy/2$ and
introducing $\chi=i\frac{8\pi G}{3\Lambda}|\alpha^2||\tau_0^2|e^{xt}\zeta$ the equations of interest can be written as
\begin{equation}
\begin{array}{l}
\displaystyle \cos(yt)\ddot{\chi}+2(\sqrt{x^2+y^2}\sin(yt+\sigma_b)-x\cos(yt))\dot\chi+{}\\
\displaystyle{} +\left(\cos(yt)\left(6h_0\sqrt{x^2+y^2}e^{x(t-t_0)}\sin(y(t-t_0)-\sigma_b)
+\frac{k^2}{a_0^2}+x^2\right)-\right.\\
\displaystyle
\left.-2x\sqrt{x^2+y^2}\sin(yt+\sigma_b)\right)\chi=-{2y}\varepsilon
\end{array}
\label{deltaijepsexMcos}
\end{equation}
\begin{equation}
\begin{array}{l}
\displaystyle \cos(yt)\ddot\varepsilon+2\sqrt{x^2+y^2}\sin(yt-\sigma_b)\dot\varepsilon+{}\\
\displaystyle{}+3H_0\sqrt{\left(\frac{k^2}{3a_0^2H_0}-x\right)^2+y^2}
\cos(yt-\sigma_c)\varepsilon=\frac{2k^2y}{a_0^2}\chi.
\end{array}
\label{deltaepsijexMcos}
\end{equation}
where all the constant coefficients are real, $\varepsilon$ and $\chi$
are real, $t_0$ is expected to be negative and
\begin{equation*}
\sigma_b=\arcsin\frac{x}{\sqrt{x^2+y^2}},\quad\sigma_c=\arcsin\frac{y}{\sqrt{(\frac{k^2}{3a_0^2H_0}-x)^2+y^2}}.
\end{equation*}

The most alarming points of the evolution are $yt=\frac{\pi}2+n\pi$
where the coefficients in front of second derivatives become zero.
Numeric integration may hit problems at these points if the precision
is not very high. In the neighborhood of these points one has
\begin{equation}
t\ddot{\chi}-2\dot\chi+2x\chi=2\varepsilon
\label{deltaijepsexMcos0}
\end{equation}
\begin{equation}
t\ddot\varepsilon-2\dot\varepsilon-3H_0\varepsilon=\frac{2k^2}{a_0^2}\chi.
\label{deltaepsijexMcos0}
\end{equation}
For negative $x$ the solution for $\varepsilon$ around $t=0$ is
$\varepsilon=\varepsilon_0+\varepsilon_1t+\dots$ meaning that these
points are not singular for the above system of equations.

A typical behavior for the function $\varepsilon$ is dumped oscillations depicted in Fig.~1.
\begin{figure}[h]
\centering
\includegraphics[width=55mm]{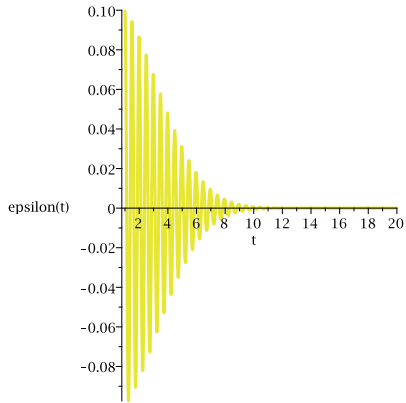}
\caption{Typical behavior of the function $\varepsilon$.}
\label{Figu1}
\end{figure}
Such a behavior does not depend on the wavenumber meaning that
perturbations with complex conjugate scalar fields do vanish. This is
different from usual models with real scalar fields where different
regimes exist and most likely growing modes are present.

Application of the curvature and entropy perturbation analysis (formulae
(\ref{isocurv}) and (\ref{entropy})) shows that both quantities decay
exponentially in the presence of a pair of complex conjugate roots. Moreover
accounting of the effect of possible other modes which correspond to other roots
with the property that these modes are decoupled from the full system of
perturbation equations (as in (\ref{deltaijepshomo})--(\ref{deltaDec})) does not
change qualitatively the result.
More comprehensive analysis on what happens when complex roots which are
significant for perturbations are present can be found in \cite{GK}.



\section{The linearized model with double roots}
Let us  consider an analytical function
$\Fc(J)$, which has simple roots $J_i$ and double roots $\tilde{J}_k$,
and the function
\begin{equation}
  \label{tau_0}
  \tau_B=\sum\limits_{i=1}^{N_1}\tau_i+\sum\limits_{k=1}^{N_2}\tilde\tau_k,
 \quad\mbox{where}\quad(\Box-J_i)\tau_i=0, \quad
(\Box-\tilde{J}_k)^2\tilde\tau_k=0.
\end{equation}
The fourth order differential equation
$(\Box-\tilde{J_k})(\Box-\tilde{J_k})\tilde\tau_k=0$ is equivalent
to the following system of the second order equations:
\begin{equation*}
  (\Box-\tilde{J_k})\tilde\tau_k=\sigma_k,\qquad
  (\Box-\tilde{J_k})\sigma_k=0.
\end{equation*}
It is convenient to write $\Box^m\tilde\tau_k$ in terms of
$\tilde\tau_k$ and $\sigma_k$:
\begin{equation}
\Box^m
\tilde\tau_k=\tilde{J}_k^m\tilde\tau_k+m\tilde{J}_k^{m-1}\sigma_k.
\label{Boxl_doubleroot}
\end{equation}
The energy--momentum tensor, which corresponds to $\tau_B$, has the
following form~\cite{Vernov2010}:
\begin{equation}
  T_{\mu\nu}\left(\tau_B\right)=
\sum\limits_{i=1}^{N_1}T_{\mu\nu}(\tau_i)+\sum\limits_{k=1}^{N_2}T_{\mu\nu}
(\tilde\tau_k),
  \label{Tmunugen}
\end{equation}
where
\begin{equation*}
  \label{TEV}
T_{\mu\nu}(\tau)=\frac{1}{g_o^2}\Bigl(E_{\mu\nu}(\tau)+E_{\nu\mu}(\tau)-g_{
\mu\nu}\left(g^{\rho\sigma}
    E_{\rho\sigma}(\tau)+W(\tau)\right)\Bigr),
\end{equation*}
\begin{equation*}
  E_{\mu\nu}(\tau_i)=\frac{{
      \Fc'(J_i)}}{2}\partial_{\mu}\tau_i\partial_{\nu}\tau_i,\qquad
  W(\tau_i)=\frac{J_i \Fc'(J_i)}{2}\tau_i^2, \label{EWsimpleroot}
\end{equation*}
\begin{equation*}
  \label{Edr} E_{\mu\nu}(\tilde\tau_k)= \frac{{
\Fc''(\tilde{J}_k)}}{4}
\left(\partial_\mu\tilde\tau_k\partial_\nu\sigma_k+\partial_\nu\tilde\tau_k
\partial_\mu\sigma_k\right)+
  \frac{\Fc'''(\tilde{J}_k)}{12}\partial_\mu\sigma_k\partial_\nu\sigma_k,
\end{equation*}
\begin{equation*}
  \label{Vdr} W(\tilde{\tau}_k)=\frac{\tilde{J}_k
    \Fc''(\tilde{J}_k)}{2}\tilde\tau_k\sigma_k+ \left(\frac{{\tilde{J}_k
        \Fc'''(\tilde{J}_k)}}{12}+\frac{{
        \Fc''(\tilde{J}_k)}}{4}\right)\sigma_k^2.
\end{equation*}
Formulae (\ref{tau_0}) and (\ref{Tmunugen}) generalize formulae (\ref{tau_sum})
and (\ref{EOJ_g_onshell}) for the functions
$\Fc(J)$ with both simple, and double roots.
Using (\ref{deltabox1}), (\ref{Boxl_doubleroot}) and
\begin{equation*}
\sum_{m=0}^{n-1}mx^{m-1}=\frac{d}{dx}\sum_{m=0}^{n-1}x^{m}=\frac{d}{dx}
\left(\frac{1-x^n}{1-x}\right)=
\frac{(n-1)x^n-nx^{n-1}+1}{(1-x)^2},
\end{equation*}
we get
\begin{equation}
\label{deltaboxDR}
\begin{array}{l}
\delta(\Box^n\tau_B)=\Box^n(\delta\!\tau_B)+\sum\limits_{m=0}^{n-1}
\Box^m(\delta\Box)\Box^{n-1-m}\tau_B={}\\[1mm]
{ \quad }
=\Box^n(\delta\!\tau_B)+\sum\limits_{i=1}^{N_1}\frac{\Box^n-J^n_i}{\Box-J_i}
(\delta\Box)\tau_i+
\sum\limits_{k=1}^{N_2}\left[\frac{\Box^n-\tilde{J}^n_k}{\Box-\tilde{J}_k}
(\delta\Box)\tilde{\tau}_k+{}\right.\\[1mm]
{ \quad } +\left.
\frac{\tilde{J}_k\left(\Box^n+(n-1)\tilde{J}_k^n-n\tilde{J}_k^{n-1}\Box\right)}{
(\Box-\tilde{J}_k)^2}(\delta\Box)\sigma_k\right].
\end{array}
\end{equation}

To formulate the perturbation equations we note that from  (\ref{Tmunugen}) one
finds
 \begin{equation*}
\varrho(\tau_B)-p(\tau_B)=2\left(\sum\limits_{i=1}^{N_1}W(\tau_i)+\sum\limits_{
k=1}^{N_2}W(\tilde{\tau}_k)\right),
 \end{equation*}
 so that,
 \begin{equation*}
 \begin{array}{l}
 \delta\varrho(\tau_B)- \delta
p(\tau_B)=2\sum\limits_{i=1}^{N_1}J_i\Fc'(J_i)\tau_i\delta\!\tau_i+ {}\\
{}+\sum\limits_{k=1}^{N_2}\left[\tilde{J}_k\Fc''(\tilde{J}
_k)(\sigma_k\delta\tilde{\tau}_k+\tilde{\tau}_k\delta\sigma_k)+
 \left(\frac{\tilde{J}_k\Fc'''(\tilde{J}_k)}{3}+\Fc''(\tilde{J}
_k)\right)\sigma_k\delta\sigma_k\right]
 \end{array}
 \end{equation*}
and
  \begin{equation}
 \begin{array}{l}
 v^s(\tau_B)=\frac k
{a(\varrho+p)}\left(\sum\limits_{i=1}^{N_1}
J_i\Fc'(J_i)\dot\tau_i\delta\!\tau_i+\right.\\
 \left.
{}+\sum\limits_{k=1}^{N_2}\left[\frac{\Fc''(\tilde{J}_k)}{2}(\dot{\sigma}
_k\delta\tilde{\tau}_k+\dot{\tilde{\tau}}_k\delta\sigma_k)+
 \frac{\Fc'''(\tilde{J}_k)}{6}\dot\sigma_k\delta\sigma_k\right]\right),
 \end{array}
 \label{vsDR}
 \end{equation}
where
 \begin{equation*}
\varrho(\tau_B)+p(\tau_B)=2\left(\sum\limits_{i=1}^{N_1}E_{00}
(\tau_i)+\sum\limits_{k=1}^{N_2}E_{00}(\tilde{\tau}_k)\right).
 \end{equation*}
Then for one double root we obtain
\begin{equation}
\begin{array}{rcl}
\Delta({\tilde{\tau}_k})&=&\frac{\Fc''(\tilde{J}_k)}{6}\Bigl\{3\tilde{J}
_k\Fc''(\tilde{J}_k)\left(\sigma_k{\dot{\tilde{\tau}}_k}^2\delta\sigma_k
+\dot\sigma_k^2\tilde{\tau}_k\delta\!\tilde{\tau}_k\right)+{}\\
&+&3\Fc''(\tilde{J}_k)\sigma_k\dot\sigma_k^2\delta\!{\tilde\tau}_k+\tilde{J}
_k\Fc'''(\tilde{J}_k)\sigma_k\dot\sigma_k\dot{\tilde\tau}_k\delta\sigma_k
\Bigr\}.
\end{array}
\end{equation}
Note that $\Delta({\tilde{\tau}_k})\neq 0$, because $\Fc''(\tilde{J}_k)\neq 0$.

After the diagonalization\footnote{Explicit formulae are given
in~\cite{Vernov2010}} of the kinetic part of the energy--momentum tensor,
one can use the general formulae for perturbations in cosmological models with
many scalar fields~\cite{hwangnoh} and get the closed
system of equations for perturbations. So, we can conclude that in the case of
the function $\Fc(\Box)$ with both simple and double roots
we get the system of local equations. We plan to consider deeper the case
of the function $\Fc(\Box)$ with double roots in our forthcoming paper.


\section{Summary and outlook}

The main results of this paper are the construction of the perturbation
equations in non-local models with one scalar field and arbitrary
potential and consideration of the very intriguing example of
perturbations. Namely, using the possibility to construct a local
equivalent model with many scalar fields we find out that masses of
these local fields may easily become complex and such a case
constitutes the above-mentioned example. The characteristic feature of
the present setup is that all the local fields in fact are not physical
and play a role of auxiliary functions introduced for the reduction of
the complicated non-local problem to a known one.
As it was noted in \cite{Koshelev07,AJV0701}, for a very wide class of
the SFT inspired models the local counterpart is not yet studied.
Looking strange such
configurations do not produce a problem for the model since they are
not physical quantities.

Perturbation equations for this local model are
(\ref{deltaGIeps01nlscalar}) and (\ref{deltaijeps}) where only $N-1$
functions $\zeta_{1j}$ are independent. The discussion on how a
cosmological constant can be generated during the tachyon evolution is
presented in~\cite{IA1,Koshelev07}. We note that perturbations in a
quintom model very close to our setup with a phantom field without
potential and an ordinary scalar field with quadratic potential were
studied in~\cite{Brandenberger08}. Perturbations in models with many
scalar fields were studied in literature considering various
cosmological scenarios \cite{hwangnoh,manyperturbations}.

In the present paper we
have worked the indicative example where two scalar fields with complex conjugate
masses are present.
We have demonstrated numerically that in the case $|\sqrt{J}|\gg H_0$
the gauge invariant energy density perturbation associated with
the matter sector does decay in all wavelength regimes in contrary to ordinary
scalar field models. The
general case of complex masses deserves deeper investigation and is partially
considered in \cite{GK}.

Moreover we singled out configurations when really an infinite set of scalar
fields may present but on the other hand there is a chance to analyze in full
the system of equations for perturbations. These are configurations when
$\sqrt{J_{i+1}}-\sqrt{J_i}\to0$ and when
$\sqrt{J_{i+1}}-\sqrt{J_i}=c_J=\const$. This should definitely help in
understanding how important quantities like curvature perturbations and entropy
perturbations behave. Also the case of double roots is addressed and it is
shown that one can again formulate the local equivalent theory and build a
closed system of equations. However, the deep analysis of all these regimes
seems to be rather involved and therefore we put it as still open question to be
addressed separately.

Looking further it is interesting to consider perturbations in
other non-local models coming from the SFT. For instance, models where
open and closed string modes are non-minimally coupled may be of
interest in cosmology. An example of the classical solution is
presented in \cite{AKd}. Furthermore it should be possible to extend
the formalism presented in this paper to other models involving
non-localities like modified gravity setups \cite{nonlocal,sftnonlocal}.


\section*{Acknowledgements}

The authors are grateful to I.Ya.~Aref'eva, B.~Craps, B.~Dragovich, and
V.F.~Mu\-kha\-nov for useful comments and discussions. This work is
supported in part by RFBR grant 08-01-00798 and state contract of
Russian Federal Agency for Science and Innovations 02.740.11.5057. A.K.
is supported in part by the Belgian Federal Science Policy Office
through the Interuniversity Attraction Poles IAP VI/11, the European
Commission FP6 RTN programme MRTN-CT-2004-005104 and by FWO-Vlaanderen
through the project G.0428.06. S.V. is supported in part by the grant
of Russian Ministry of Education and Science NSh-4142.2010.2 and in part by CPAN10-PD12 (ICE, Barcelona, Spain).

\end{document}